\documentclass[aps,pra,twocolumn,superscriptaddress,longbibliography]{revtex4-1}

\usepackage{natbib}
\usepackage{pgfplots}
\usepackage{amssymb}
\usepackage{amsmath,amsthm}
\usepackage{amsfonts,dsfont}
\usepackage{graphicx}
\usepackage{mathtools}
\usepackage{tikz}
\usetikzlibrary{positioning}
\usepackage[normalem]{ulem}
\usepackage{color}
\usepackage{xcolor}

\newcommand{\bd}[1]{\boldsymbol{#1}}

\newcommand{\g}{\gamma}

\newcommand{\G}{\Gamma}
\newcommand{\Fp}{\mathcal{F}^{(p)}}
\newcommand{\Fe}{\mathcal{F}^{(e)}}
\newcommand{\FN}{\mathcal{F}_N}
\newcommand{\F}{\mathcal{F}}

\newcommand{\FpN}{\mathcal{F}^{\,(p)}_N}
\newcommand{\Fpd}{\mathcal{F}^{\,(p)}_2}

\newcommand{\FeN}{\mathcal{F}^{\,(e)}_N}

\newcommand{\bra}[1]{\mbox{$\langle #1 |$}}
\newcommand{\ket}[1]{\mbox{$| #1 \rangle$}}

\begin{document}
\title{Reduced Density Matrix Functional Theory for Bosons}

\author{Carlos L. Benavides-Riveros}
\affiliation{Institut f\"ur Physik, Martin-Luther-Universit\"at
Halle-Wittenberg, 06120 Halle (Saale), Germany}
\affiliation{NR-ISM, Division of Ultrafast Processes in Materials (FLASHit), Area della Ricerca di Roma 1, Via Salaria Km 29.3, I-00016 Monterotondo Scalo, Italy}

\author{Jakob Wolff}
\affiliation{Institut f\"ur Physik, Martin-Luther-Universit\"at
Halle-Wittenberg, 06120 Halle (Saale), Germany}

\author{Miguel A. L. Marques}
\affiliation{Institut f\"ur Physik, Martin-Luther-Universit\"at
Halle-Wittenberg, 06120 Halle (Saale), Germany}

\author{Christian Schilling}
\email{c.schilling@physik.uni-muenchen.de}
\affiliation{Department of Physics, Arnold Sommerfeld Center for Theoretical Physics,
Ludwig-Maximilians-Universit\"at M\"unchen, Theresienstrasse 37, 80333 M\" unchen, Germany}
\affiliation{Wolfson College, University of Oxford, Linton Rd, Oxford OX2 6UD, United Kingdom}

\date{\today}

\begin{abstract}
Based on a generalization of Hohenberg-Kohn's theorem, we propose a ground state theory for bosonic quantum systems.
Since it involves the one-particle reduced density matrix $\g$ as a natural variable but still recovers quantum correlations in an exact way it is particularly well-suited for the accurate description of Bose-Einstein condensates.
As a proof of principle we study the building block of optical lattices. The solution of the underlying $v$-representability problem is found and its peculiar form identifies the constrained search formalism as the ideal starting point for constructing accurate functional approximations: The exact functionals for this $N$-boson Hubbard dimer and general Bogoliubov-approximated systems are determined. The respective   gradient forces are found to diverge in the regime of Bose-Einstein condensation, $\nabla_\g \F \propto 1/\sqrt{1-N_{\mathrm{BEC}}/N}$, providing a natural explanation for the absence of complete BEC in nature.
\end{abstract}

\maketitle

\paragraph*{Introduction.---}
One of the striking features of quantum many-body systems is that their particles interact only by two-body forces. As a consequence, the ground state problem can in principle be solved in terms of the two-particle reduced density matrix replacing the exponentially complex $N$-particle wave function \cite{Col63,Percus64,Mazz12,Mazz16}. Furthermore, in each scientific field all systems of interest are characterized by the same \emph{fixed} interaction $W$ between the particles (e.g., Coulomb interaction in quantum chemistry and  contact interaction in the field of ultracold gases). The class of $N$-particle Hamiltonians is thus parameterized solely by the external potential $v$.
Since the conjugate variable of $v$ is the particle density, this heuristic reasoning identifies density functional theory as the most economic approach in each scientific field for addressing the ground state problem. %for various systems of interest in each given field of physics.
Indeed, density functional theory has become in the past few decades the method of choice for electronic structure calculations in physics, chemistry and materials science \cite{Jones15}.
It is, however, not suitable for describing in a direct way Bose-Einstein condensation (BEC),
one of the most fascinating phenomena of quantum physics. This is due to the fact that
the particle density does in general not provide sufficient insights into the presence or absence of BEC, quite in contrast to the one-particle reduced density matrix (1RDM)
\begin{equation}\label{1RDM}
\g \equiv N \mbox{Tr}_{N-1}[\G]\equiv \sum_{\alpha} n_\alpha \ket{\alpha}\!\bra{\alpha}\,,
\end{equation}
which is obtained from the $N$-boson density operator $\G$ by integrating out all except one boson:
BEC is present whenever the largest eigenvalue $n_{max} = \max_{\varphi}\bra{\varphi}\g\ket{\varphi}$ of the 1RDM is proportional to $N$ \cite{Penrose1956}. This criterion is more general than the one referring to off-diagonal long-range order of $\gamma(\vec{r},\vec{r}\,')= \bra{\vec{r}}\g\ket{\vec{r}\,'}$ \cite{Yang1962}, since it also applies to non-homogeneous systems.

While the theoretical prediction of BEC \cite{Bose,Einstein} traces back to the 1920s,
its experimental realization for atoms in 1995 \cite{Anderson1995,Ketterle1995,Bradley1995} has led to a renewed interest. Since then, the res\-pec\-tive
field of ultracold atomic gases has even beco\-me one of the most active fields in quantum physics (see, e.g., Refs.~\cite{Dalfovo1999,Bloch2008,Chin2010,Weidemueller2011}) with a broad range of applications in quantum technologies (see, e.g., Refs. \cite{Fadel409,Kunkel413,Lange416, Schmied441}). It is also this development which urges us to
propose and work out in the following a computationally feasible method which is capable of describing strongly interacting bosons in general and BEC in particular.
This bosonic one-particle reduced density matrix functional theory (RDMFT) is based
on a generalization of the famous Hohenberg-Kohn theorem \cite{Gilbert}. It therefore recovers quantum correlations in an effective but exact manner and
is not restricted to the low-density regime, quite in contrast to the Gross-Pitaevskii
\cite{Pitaevskii1961,Gross1963,pita}
or Bogoliubov theory \cite{Bogo,pita}. The study of two concrete systems shall serve as a proof of principle: We succeed in determining their universal functionals and solve the underlying $v$- and $N$-representability problem which have partly hampered the development of RDMFT in fermionic quantum systems. A natural explanation for the absence of complete BEC in nature follows, highlighting the potential of our novel method.

\paragraph*{Foundation of bosonic RDMFT.---}

Due to Gilbert \cite{Gilbert}, a generalization of the Kohn-Hohenberg theorem to Hamiltonians
\begin{equation}\label{ham}
H(h) \equiv h+W
\end{equation}
with a fixed interaction $W$ proves the existence of a universal 1RDM-functional $\mathcal{F}[\g]$: The ground state energy and ground state 1RDM
follow for any choice of the one-particle Hamiltonian $h$ from the minimization of the total energy functional
\begin{equation}\label{energy-h}
\mathcal{E}_h[\g]=  \mbox{Tr}[h \g] +
\mathcal{F}[\g]  \,.
\end{equation}
The functional $\mathcal{F}$ is universal in the sense that it does not depend on $h\equiv t+v$ but only on the fixed interaction $W$. This is due to the fact that the 1RDM $\g$ allows one to determine not only the external potential energy, $\mbox{Tr}[v \g]$, but also the kinetic energy, $\mbox{Tr}[t \g]$. Due to the significance of bosonic quantum systems and the importance of $\g$ as an indicator for BEC it is surprising that RDMFT has been developed only for \textit{fermionic} systems (see, e.g., the reviews \cite{C00,M07,PG16,SKB17}). In the following we take the first steps towards realizing a bosonic RDMFT and in particular observe that some obstacles in case of fermionic systems do not hamper its bosonic counterpart.

Let us first recall that the universal functional $\mathcal{F}$ is defined on the set $\mathcal{P}_v$ of 1RDMs which correspond to ground states of Hamiltonians $H(h)$.
But for which $\g$ does there exist a corresponding $h$? Unfortunately, no solution to this so-called $v$-representability problem is known, neither for fermions nor for bosons. To circumvent the $v$-representability problem, Levy suggested an extension of RDMFT to including nonphysical 1RDMs as well \cite{Le79} (see also Ref.~\cite{L83}). Expressing the ground state energy as $E(h)\equiv \min_{\G}\mbox{Tr}_N[H(h)\G ]$, and using the fact that the expectation value of $h$ is determined by $\g$, allows one to replace $\mathcal{F}$ in \eqref{energy-h} by \cite{Le79,V80,L83,GR19}:
\begin{equation}\label{levy}
 \mathcal{F}^{(p/e)}[\g]  = \min_{\G\mapsto \g} \mbox{Tr}_N[W\G ]\,.
\end{equation}
The minimization in \eqref{levy} may either be restricted to the \emph{pure} $(p)$ or all ensemble $(e)$ $N$-boson states $\G$ mapping to the given 1RDM, $\g=N \mbox{Tr}_{N-1}[\G]$. Consequently the functional $\mathcal{F}^{(p/e)}$ is defined on the domain $\mathcal{P}_{p/e}$ of pure/ensemble $N$-representable 1RDMs, where $\mathcal{P}_v \subseteq \mathcal{P}_p \subseteq \mathcal{P}_e$. A far-reaching observation is that for \emph{every} 1RDM (recall \eqref{1RDM}) there exists a corresponding bosonic pure state $\G\equiv \ket{\Phi}\!\bra{\Phi}$, e.g., $\ket{\Phi} = 1/\sqrt{N}\sum_{\alpha}\sqrt{n_{\alpha}}\ket{\alpha,\ldots,\alpha}$. Hence, in contrast to fermions \cite{Col63,KL06,AK08,Schilling2018}, the one-body pure $N$-representability problem is trivial. Consequently, it will not hamper the development of bosonic functionals, and one has in particular $\mathcal{P}_p = \mathcal{P}_e$.

\paragraph*{Hubbard dimer.---}
To illustrate the potential of bosonic RDMFT we discuss as a first example
the Hubbard dimer for an arbitrary number $N$ of spinless bosons.
This building block of the Bose-Hubbard model is realized \cite{Zuern2012} and prominently used
in the context of ultracold bosonic atoms, whose parameters can be tuned by laser light
\cite{PhysRevLett.81.3108,Bloch2008,Chin2010,Weidemueller2011}. Similarly to the two-electron Hubbard
dimer in the context of fermionic functional theories \cite{Pastor2011a,Pastor2011b,Wagner2012,Carrascal2015,Kamil2016,Schmidt2019}, its bosonic counterpart will serve as a theoretical laboratory system, eventually providing crucial insights into larger systems.
Its Hamiltonian reads
\begin{equation}\label{dimer}
H = -t (b_L^\dagger b_R+b_R^\dagger b_L)+\! \sum_{j=L/R}\! v_j \hat{n}_j +U\!\sum_{j=L/R}\! \hat{n}_j(\hat{n}_j-1)\,,
\end{equation}
where the operators $b^\dagger_{j}$ and $b_{j}$ create and annihilate a boson on site $j=L/R$, and
$\hat n_{j}$ is the corresponding particle-number ope\-ra\-tor. The first term in Eq.~\eqref{dimer}
describes the hopping between both sites while the second one represents
the external potential and the third one the on-site repulsion ($U >0$).

\begin{figure}[t!]
  \includegraphics[scale=0.40]{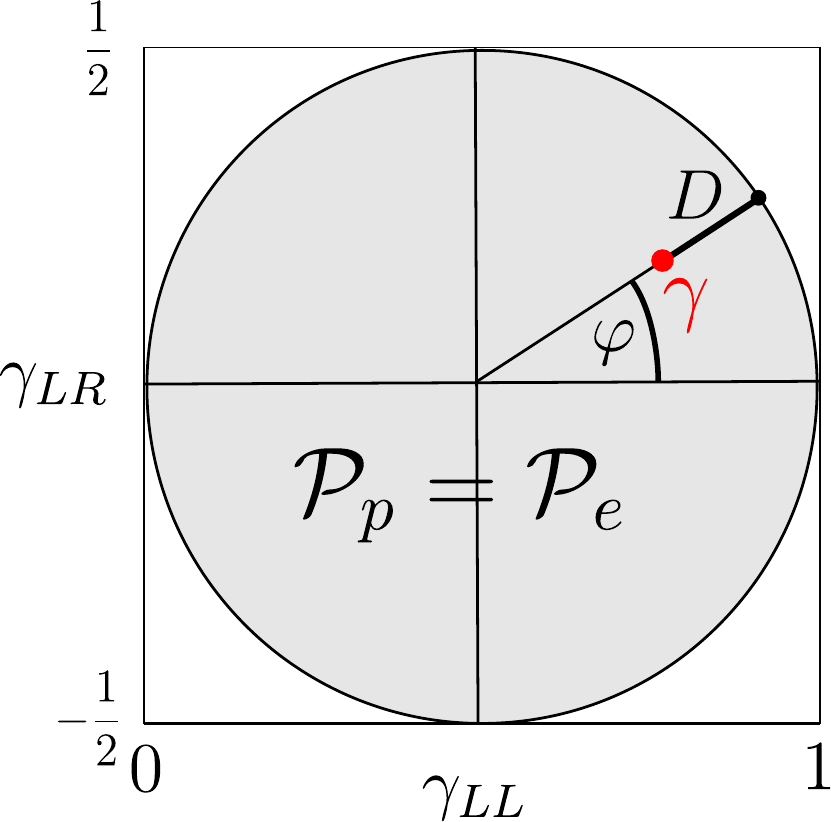}\hspace{1.0cm}
  \includegraphics[scale=0.38]{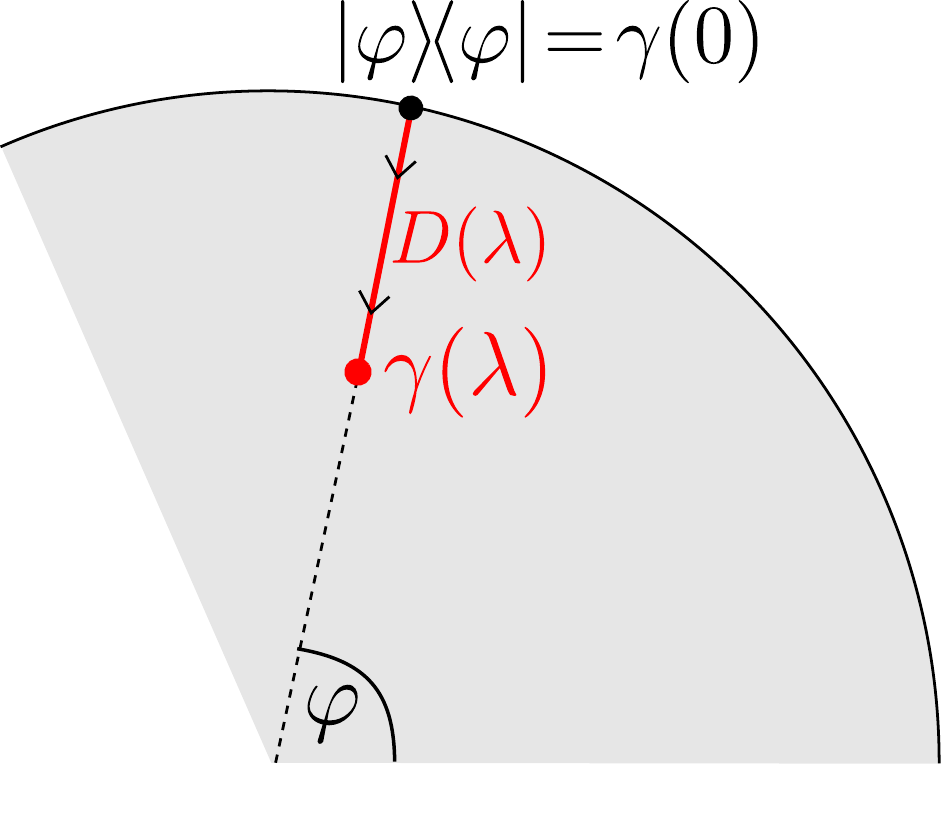}
  \caption{Left: Illustration of the spherical representation \eqref{1RDMsph} of the 1RDM $\gamma$. Right: Straight line $\g(\lambda)$ as constructed within the approach \eqref{seq} at an angle $\varphi$.}
  \label{fig:Dphi}
\end{figure}

In the following we represent $\g$ with respect to the lattice site states $\ket{L},\ket{R}$ and assume real-valued matrix elements. We choose $\g_{LL}=1-\g_{RR}$ and $\g_{LR}=\g_{RL}$ as the two independent variables.
%lattice site representation the 1RDM follows as
%\begin{equation}\label{1RDM-site}
%\gamma_{ij}= \mbox{Tr}\big[b_i^\dagger b_j \Gamma  \big]\ ,  \  i,j \in \{L,R\}  \ .
%\end{equation}
%%The phases of the one-particle basis can be chosen such that $\gamma_{LR}=\gamma_{RL}$.
Here, we normalize the 1RDM to unity since then the sets $\mathcal{P}_p =  \mathcal{P}_e$ become independent of $N$ (which allows the comparison of functionals for different values of $N$). As already stressed, the only constraint on those sets is that $\g$'s eigenvalues are nonnegative, leading to
\begin{equation}\label{N-represent}
 \gamma_{LR}^2 + (\gamma_{LL} - \tfrac12)^2 \leq \tfrac14 \,.
\end{equation}
Due to the circular symmetry of this disc it will prove convenient below to also introduce spherical coordinates:  $\gamma_{LL}=\frac{1}{2}[1+ (1-2D)\cos{\varphi}]$ and
$\gamma_{LR}=\frac{1}{2}(1-2D)\sin{\varphi}$. Hence, as illustrated in Figure \ref{fig:Dphi}, $D$ is $\g$'s distance to the boundary  $\partial\mathcal{P}_p$ and (\ref{N-represent}) reduces to $0 \leq D \leq 1/2$.
The corresponding spectral decomposition of $\g$ becomes
\begin{equation}\label{1RDMsph}
 \g(D,\varphi) = (1-D)   \ket{\varphi}\!\bra{\varphi} + D \ket{\varphi^{\perp}}\!\bra{ \varphi^{\perp}} \,,
\end{equation}
with the natural orbitals $ \ket{\varphi}= \cos( \varphi/2) \ket{L}+ \sin( \varphi/2) \ket{R}$ and $ \ket{\varphi^{\perp}}= \sin( \varphi/2) \ket{L} - \cos(\varphi/2) \ket{R}$.
%The corresponding Bose operators are denoted by $b^{\dagger}_{\varphi},b_{\varphi}$ and
%$b^{\dagger}_{\varphi^{\perp}},b_{\varphi^{\perp}}$.

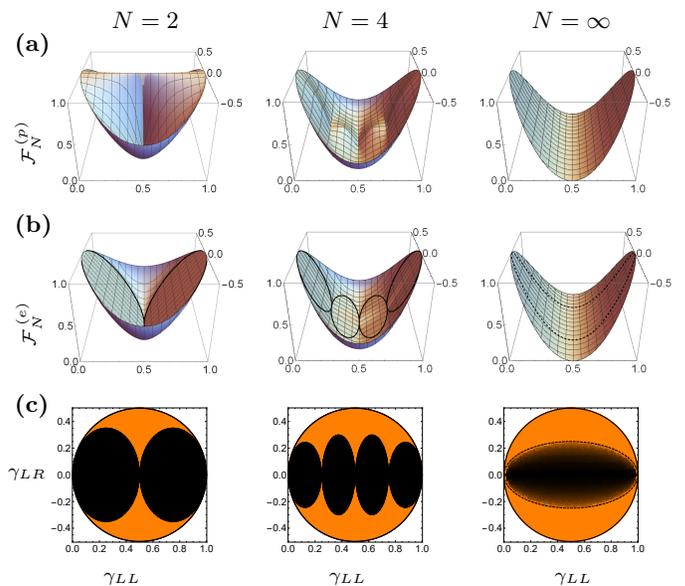
\begin{figure}[htb]
\begin{tikzpicture}
 \node (img) {\includegraphics[scale=0.25]{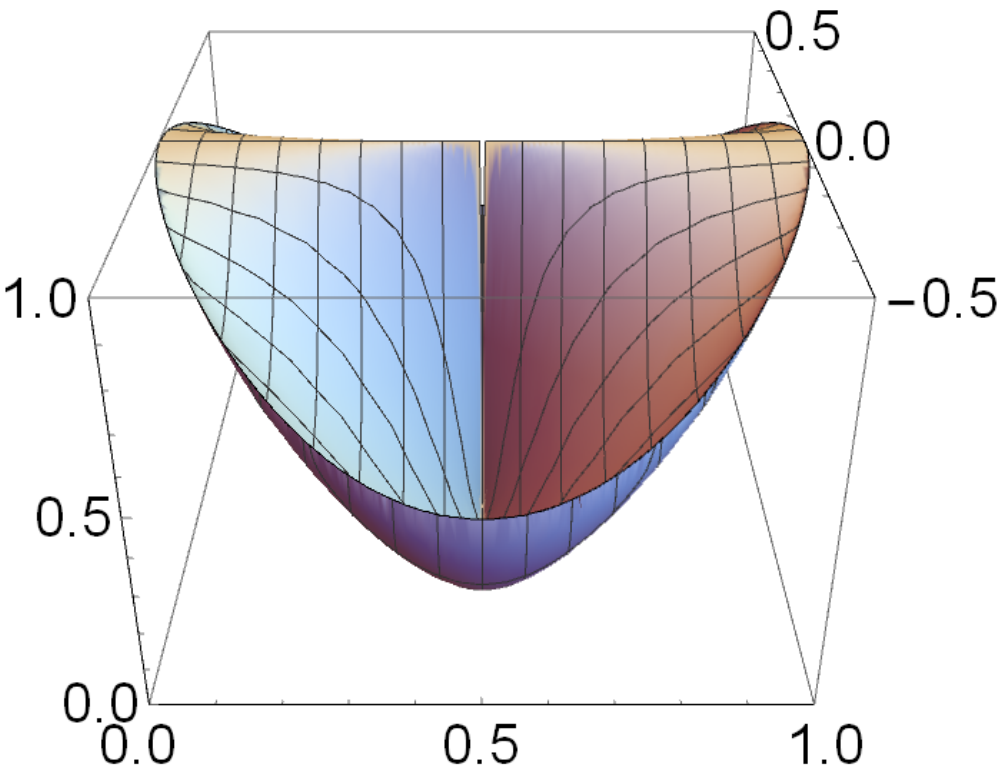}\hspace{0.2cm}
  \includegraphics[scale=0.25]{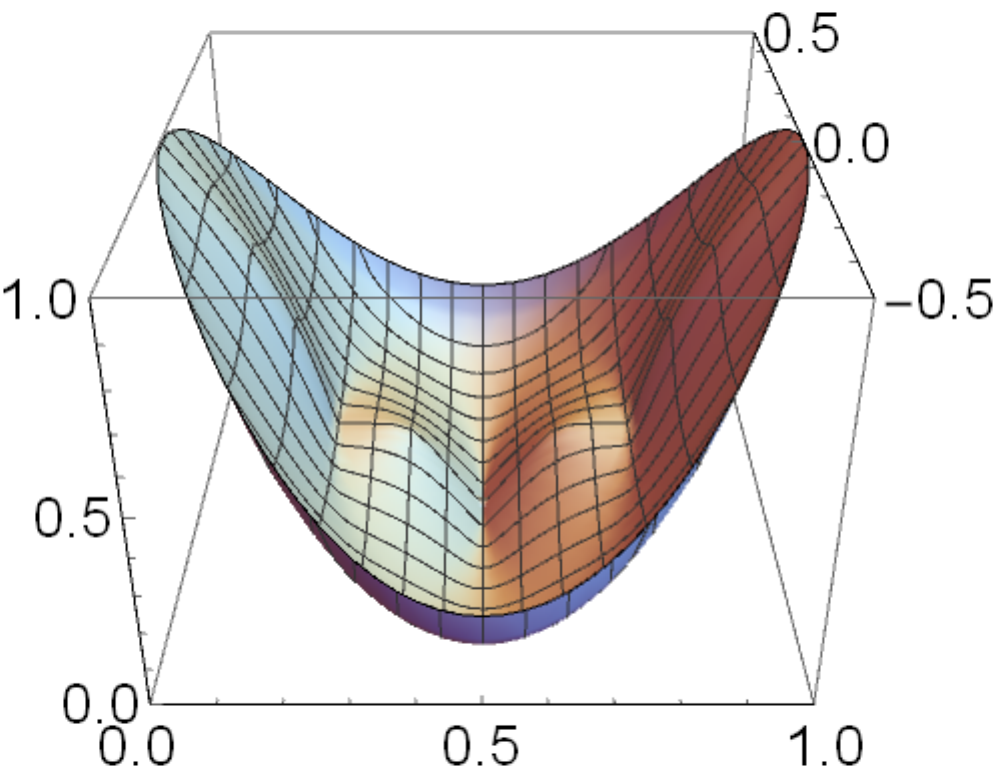}\hspace{0.2cm}
  \includegraphics[scale=0.25]{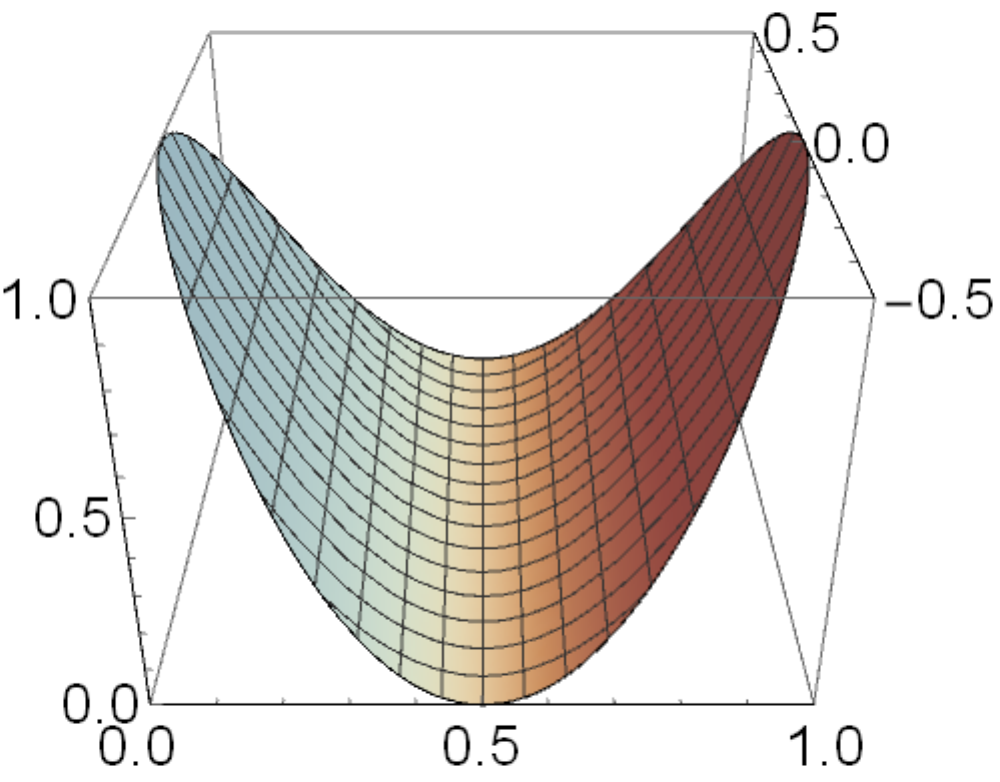}};
  \node[above=of img, node distance=0cm, xshift=0.0cm, yshift=-1.0cm,font=\color{black}] {$N=2$\hspace{1.9cm}$N=4$\hspace{1.9cm}$N=\infty$};
%  \node[left=of img, node distance=0cm, anchor=center, xshift=3.0cm,yshift=-1.8cm,font=\color{black}] {\scriptsize$\g_{LL}$};
  \node[left=of img, node distance=0cm, anchor=center, xshift=0.8cm,yshift=1cm,font=\color{black}] {\textbf{ (a)}};
  \node[left=of img, node distance=0cm, anchor=center, xshift=0.8cm,yshift=-0.3cm,font=\color{black}] {\rotatebox{90}{\scriptsize$\FpN$}};
%     \node[below=of img, node distance=0cm, xshift=2.1cm, yshift=1.2cm,font=\color{black}]{$\g_{LL}$};
%  \node[left=of img, node distance=0cm, anchor=center, xshift=8.8cm,yshift=0.8cm,font=\color{black}] {\scriptsize$\g_{LR}$};
%   \node[left=of img, node distance=0cm, anchor=center, xshift=5.1cm,yshift=-0.6cm,font=\color{black}] {\scriptsize$\mathcal{F}^{(p)}_3$};
 \end{tikzpicture}
\begin{tikzpicture}
 \node (img) {\includegraphics[scale=0.25]{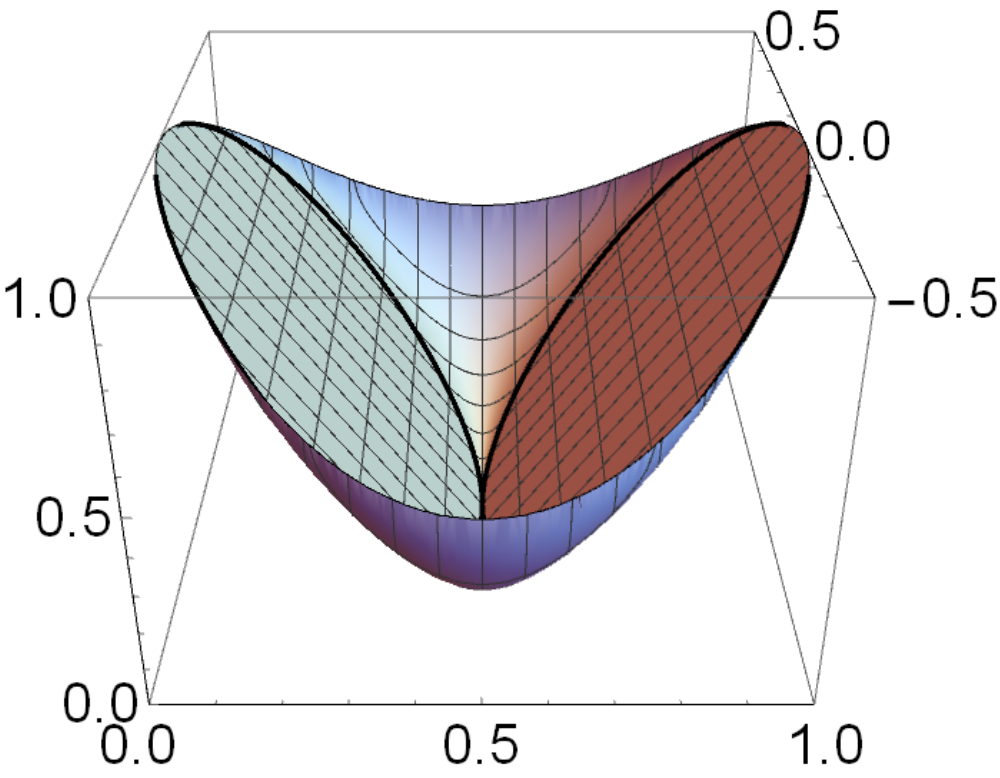}\hspace{0.2cm}
  \includegraphics[scale=0.25]{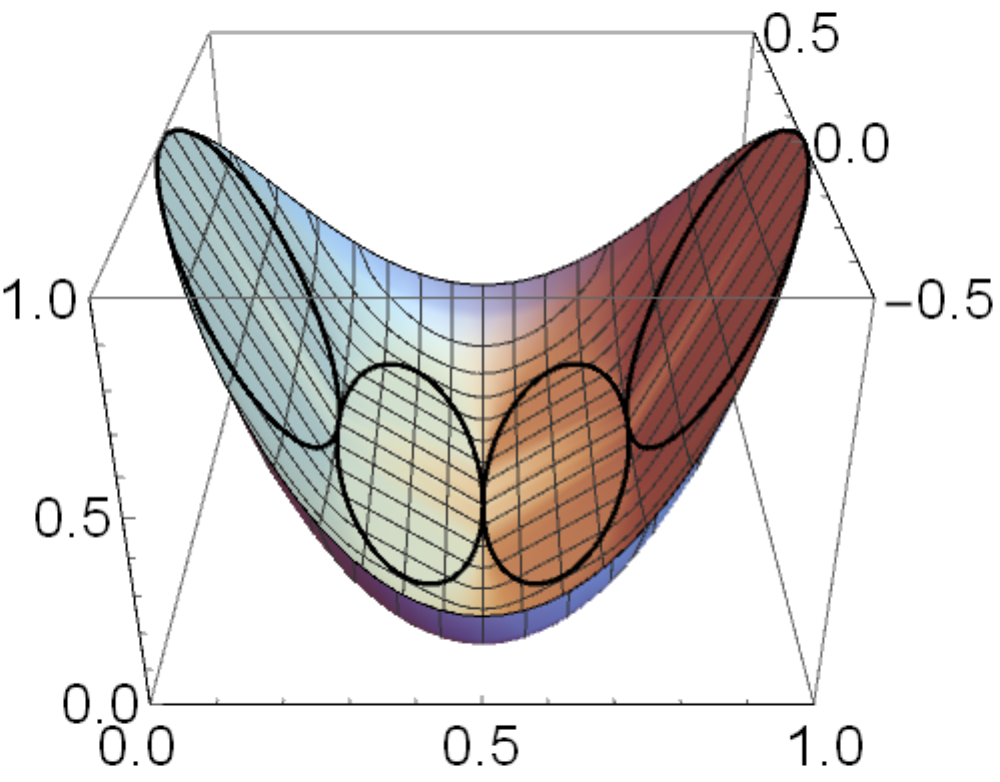}\hspace{0.2cm}
  \includegraphics[scale=0.25]{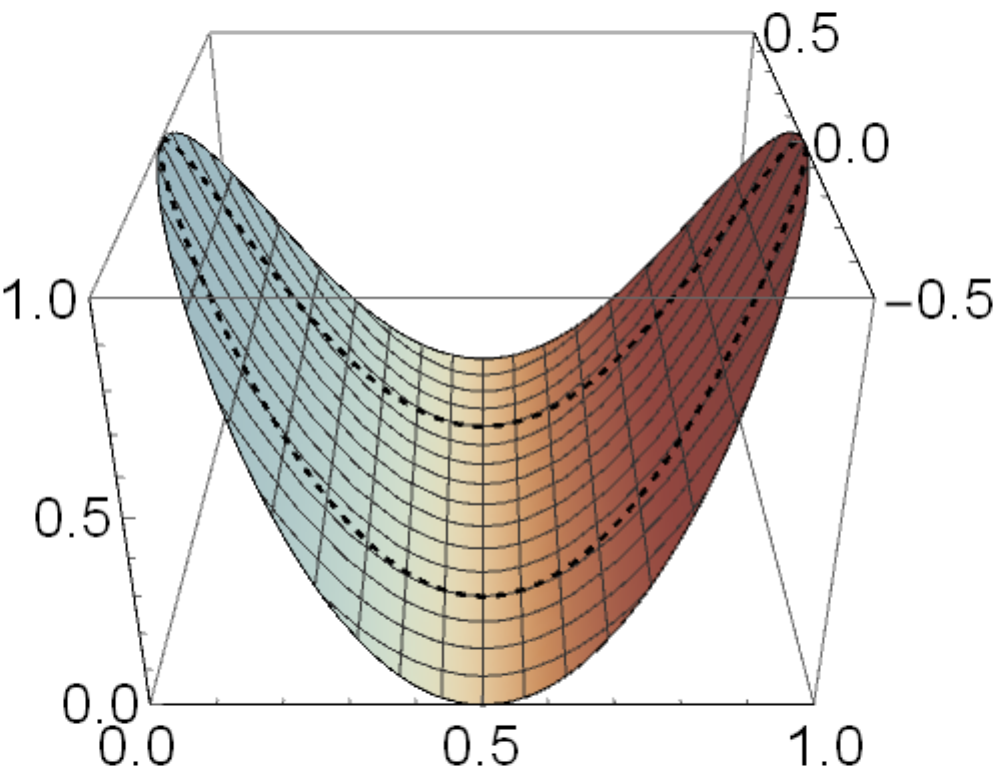}};
  \node[left=of img, node distance=0cm, anchor=center, xshift=0.8cm,yshift=1cm,font=\color{black}] {\textbf{ (b)}};
  \node[left=of img, node distance=0cm, anchor=center, xshift=0.8cm,yshift=-0.3cm,font=\color{black}] {\rotatebox{90}{\scriptsize$\FeN$}};
 \end{tikzpicture}
\begin{tikzpicture}
 \node (img) {
\includegraphics[scale=0.17]{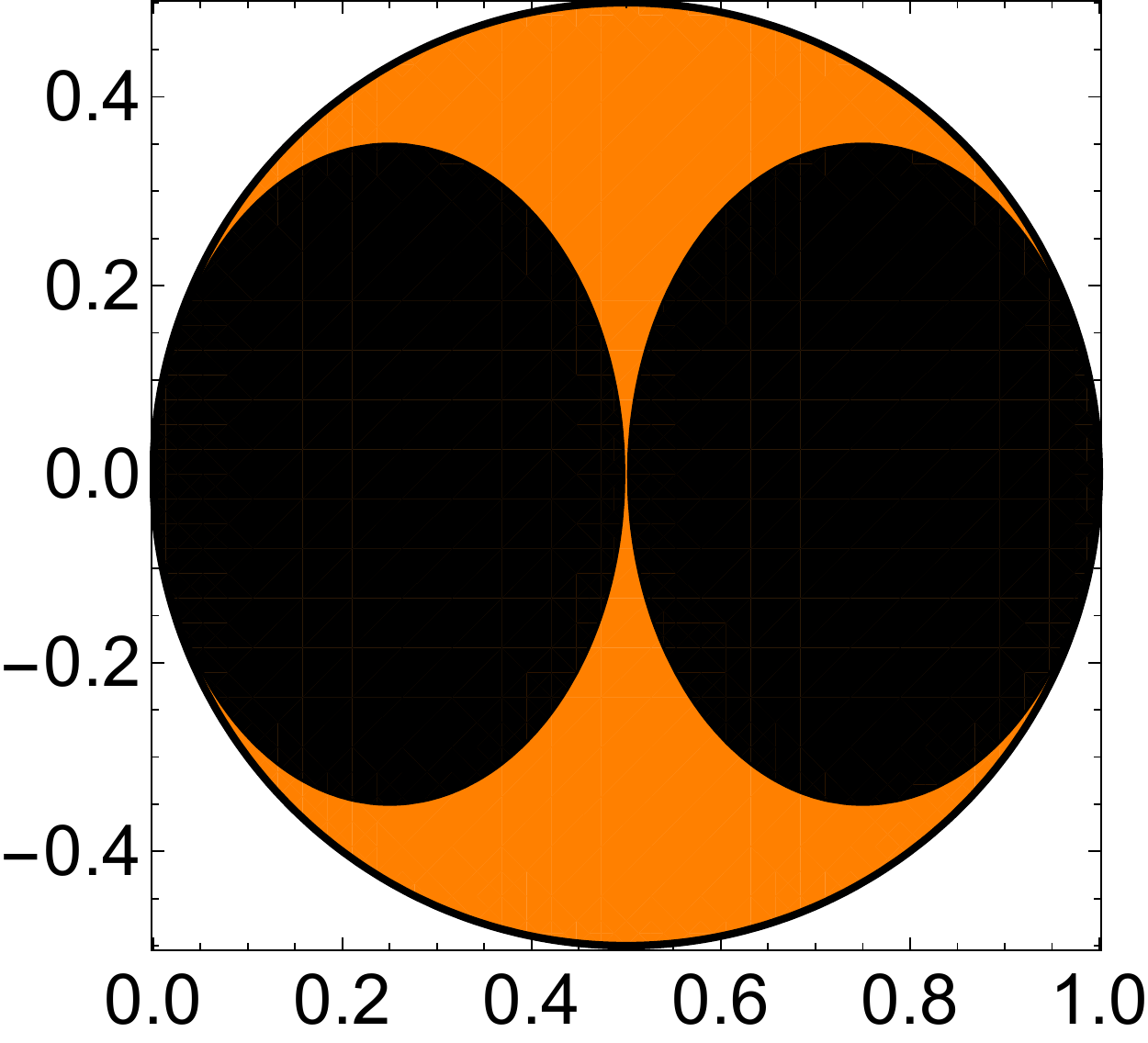}\hspace{0.6cm}
\includegraphics[scale=0.17]{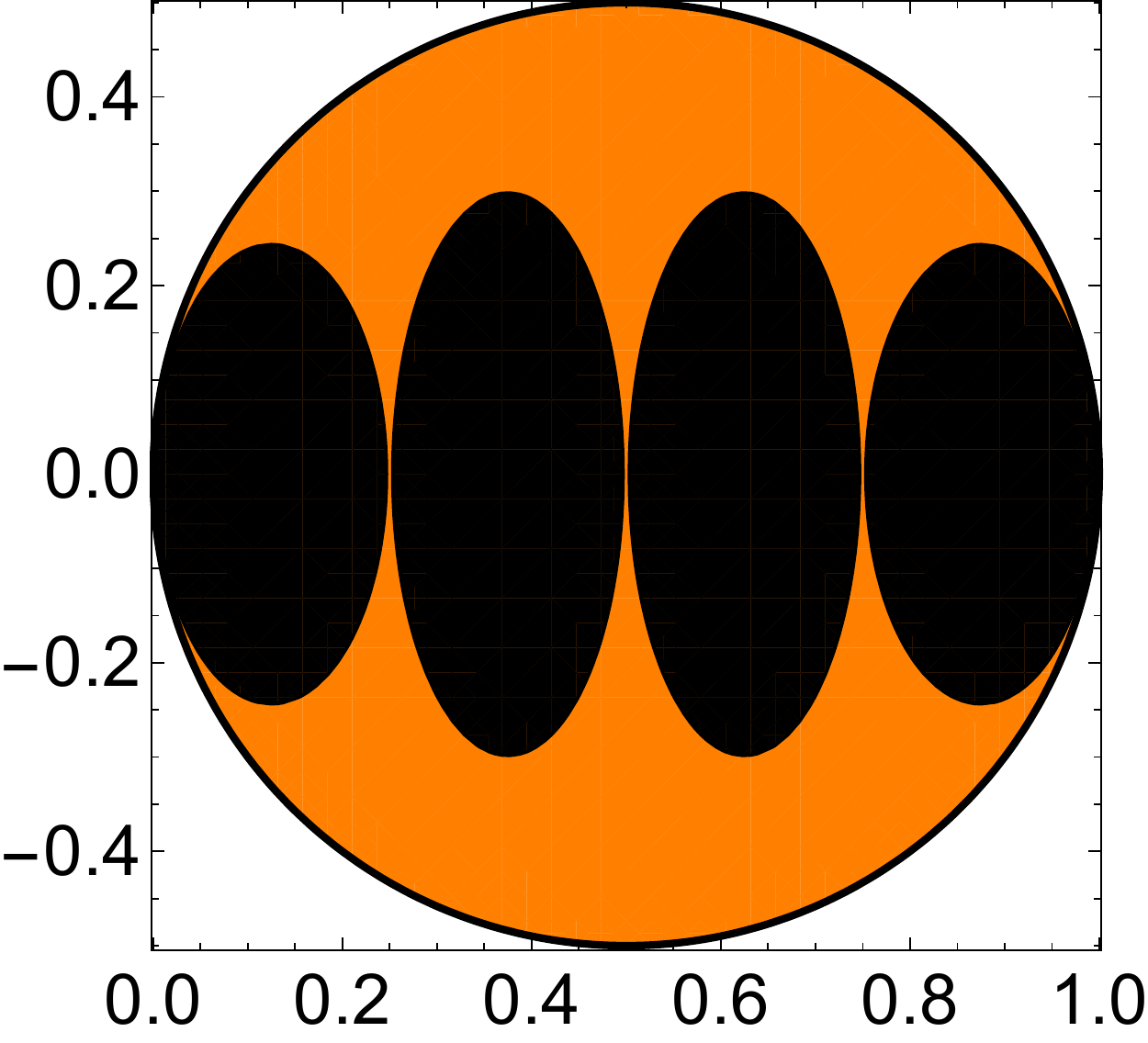}\hspace{0.6cm}
\includegraphics[scale=0.17]{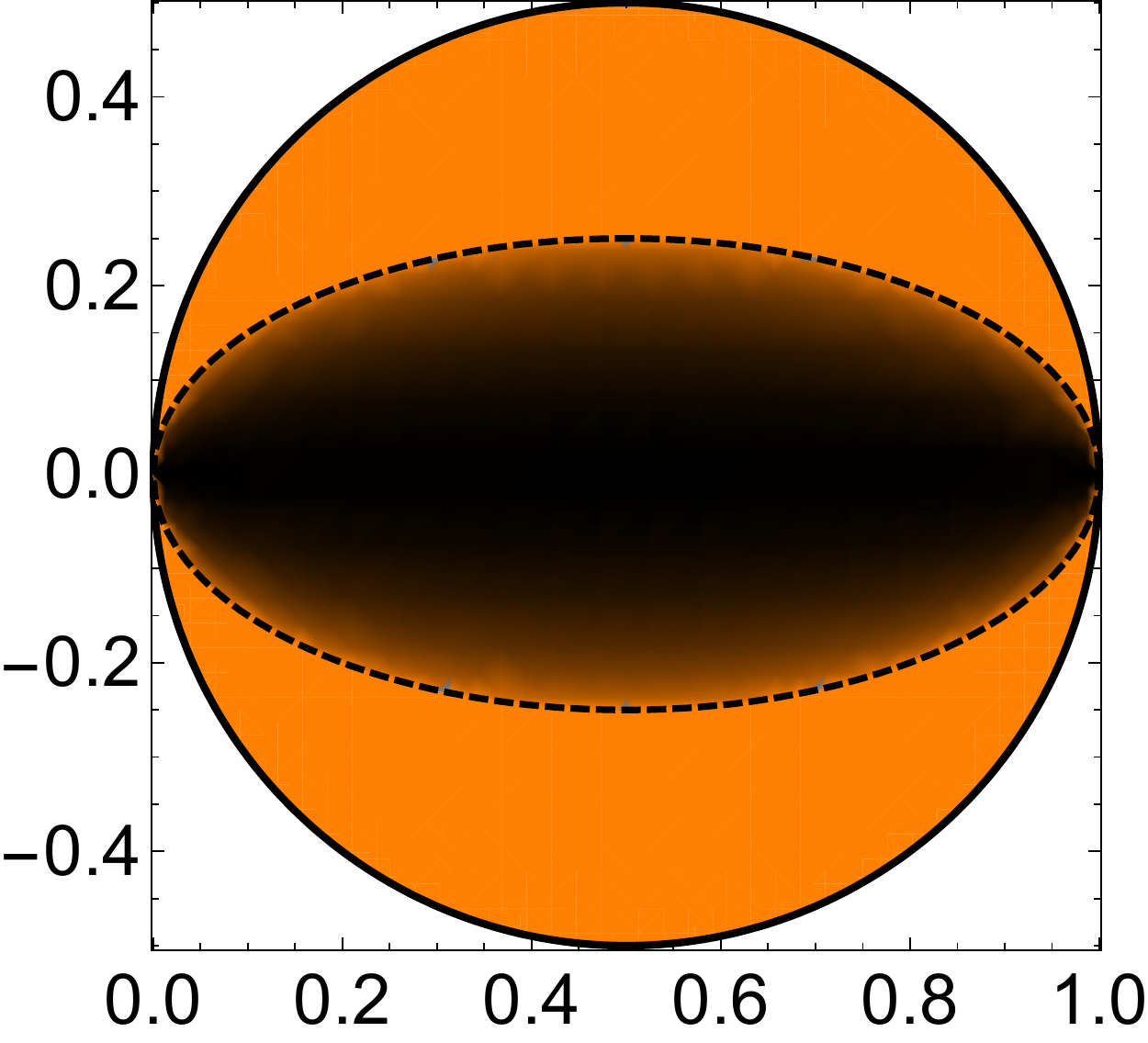}};
  \node[left=of img, node distance=0cm, anchor=center, xshift=0.8cm,yshift=0.1cm,font=\color{black}] {\scriptsize$\g_{LR}$};
  \node[left=of img, node distance=0cm, anchor=center, xshift=0.8cm,yshift=1cm,font=\color{black}] {\textbf{ (c)}};
\node[below=of img, node distance=0cm, xshift=0.0cm, yshift=1cm,font=\color{black}] {\scriptsize$\g_{LL}$};
\node[below=of img, node distance=0cm, xshift=-3.0cm, yshift=1cm,font=\color{black}] {\scriptsize$\g_{LL}$};
\node[below=of img, node distance=0cm, xshift=3.0cm, yshift=1cm,font=\color{black}] {\scriptsize$\g_{LL}$};
\end{tikzpicture}
\caption{For the Bose-Hubbard dimer we plot \textbf {(a)} the pure functional
$\FpN$ and \textbf{(b)} the ensemble functional $\FeN$ (both renormalized to $[0,1]$) as functions of the
diagonal $\gamma_{LL}$ and the off-diagonal entry $\gamma_{LR}$ of the 1RDM for the
particle numbers $N = 2,4, \infty$. In \textbf{(c)} the $v$-representable 1RDMs are shown in orange and the
nonphysical ones in black (see text for more details).}
\label{fig1}
\end{figure}

To discuss and compare the three universal functionals $\F$, $\Fp$ and $\Fe$, respectively, we first need to address the underlying $v$-representability problem. Given its fundamental significance in functional theories, it is remarkable that
no solution is known so far beyond the two-electron Hubbard dimer \cite{Cohen1}. To solve here that problem for arbitrary particle numbers $N$, we first observe that the ground states for the hopping rate $t=0$ are given by configuration states with $n_L$ bosons on the left and $n_R=N-n_L$ on the right site.
By varying the asymmetry $v_L-v_R$ of the external potential we can reach all values $n_L=0,1,\cdots,N$ and therefore each
$\g=\frac{n_L}{N}\ket{L}\!\bra{L}+ \frac{N-n_L}{N}\ket{R}\!\bra{R}$ is $v$-representable. Moreover, $v_L-v_R$ can be chosen such that the two configurations $n_L$ and $n_L+1$ are degenerate. By considering infinitesimal deformations of the respective Hamiltonian, one can thus reach any possible superposition $x \ket{n_L,n_R}\pm \sqrt{1-x^2}\ket{n_L\!+\!1,n_R\!-\!1}$. This leads to ellipses of $v$-representable 1RDMs.  As it is shown in Appendix \ref{sec:nonV}, the degeneracy of those specific ground states implies that all 1RDMs surrounded by such an ellipse (black filled ellipses in Figure \ref{fig1}) are not $v$-representable. Moreover, by anticipating the results on the presence of a diverging gradient, none of the 1RDMs on the boundary $\partial\mathcal{P}_p$ is $v$-representable (except $\g=\ket{L}\!\bra{L},\ket{R}\!\bra{R}$) but all points in its vicinity ($0 < D \ll 1$) can be obtained as ground state 1RDMs. Last but not least, each 1RDM between the boundaries of the black filled ellipses and $\partial\mathcal{P}_p$ can be reached. This can be confirmed by numerical investigations or mathematically by constructing corresponding connecting paths of ground state 1RDMs.

The solution of the $v$-representability problem provides additional crucial insights. In particular, the probability $p_N=1-\mbox{Vol}(\mathcal{P}_v)/\mbox{Vol}(\mathcal{P}_p)$ for finding non-$v$-re\-pre\-sentable 1RDMs does not vanish for large particle numbers $N$, $p_N \rightarrow \pi/8 \simeq 0.39$. Moreover, the domain $\mathcal{P}_v$ (orange) of the Gilbert functional $\F$ is getting arbitrarily complicated for larger $N$, as sketched in Figure \ref{fig1}. This identifies Levy's constrained search \eqref{levy} as the more suitable starting point for developing an RDMFT. The corresponding functional $\FpN$ can be determined analytically for $N=2$ bosons,
$\Fpd[\g(D,\varphi)]= U \left[ 2 - \left(1 + 2\sqrt{D(1-D) }\right) \sin^2(\varphi)\right]$, and in the limit of large $N$ (see Appendix \ref{sec:Fdimer}). For finite $N>2$, one can easily determine the functional by an exact numerical calculation based on the minimization in \eqref{levy}. The corresponding ensemble functionals follow directly as the lower convex envelops, $\FeN= \mbox{Conv}(\FpN)$ \cite{Schilling2018}.

The results for $\FpN$ and  $\FeN$ together with the solution of the $v$-representability problem are presented in Figure \ref{fig1}. Panel (b) confirms that  $\FeN$ is indeed given as the largest convex function fulfilling $\FeN \leq \FpN$ on the entire domain $\mathcal{P}_p$. While for $v$-representable 1RDMs, $\g \in \mathcal{P}_v$, both functionals $\FpN, \FeN$ necessarily coincide \cite{Cohen1,Cohen2,Schilling2018,Gritsenko2019} (they are equal to $\F$), this is remarkably also the case in the limit of large $N$ for
non-$v$-representable 1RDMs. %If the hold for general fermionic systems in the macroscopic limit, this would have tremendous implications: imply %that
The most surprising insight, however, is that the gradients of the functionals seem to diverge repulsively on the boundary $\partial\mathcal{P}_p$ which prevents the system from ever reaching \emph{complete} condensation in any state $\ket{\varphi}=b^{\dagger}_{\varphi} \ket{0}$ (as corresponding to $D=0$).
%, independent of the one-particle Hamiltonian $h$.
For instance, for $N=2$ one finds $\partial\Fpd\!/\partial D \simeq -U\sin^2(\varphi)/\sqrt{D}$.
Does this result generalize to larger systems and in that sense provide a natural explanation for the absence of \emph{complete} BEC in nature (quantum depletion)?

In the following we confirm the existence of this `Bose-Einstein condensation-force' also for $N>2$. For this, we propose and work out an approach which allows one to determine exact functionals in the vicinity of the boundary $\partial\mathcal{P}_p$ which corresponds to $N$-boson states close to complete BEC. We first observe that the value $\Fp[\gamma_h]$ of the functional at a $v$-representable `point' $\g_h$  (with corresponding Hamiltonian $H(h)$ and ground state energy $E(h)$)
%\begin{equation}
%h \mapsto H(h) \mapsto \ket{\Phi} \mapsto \g
%\end{equation}
follows directly from the energy relation
\begin{equation}\label{gh}
E(h)= \Fp[\g_h]+\mbox{Tr}[h \g_h]\,.
\end{equation}
The second crucial ingredient is that each $\g= \ket{\varphi}\!\bra{\varphi}\in \partial\mathcal{P}_p$ has a unique corresponding $N$-boson state which is given by $1/\sqrt{N!}\,b_{\varphi}^N\ket{0}$. We could identify those states as the unique ground states of the one-particle Hamiltonians $h^{(0)}\equiv -b_{\varphi}^\dagger b_{\varphi}$. As illustrated in Figure \ref{fig:Dphi}, the idea is then to construct for fixed $\varphi$ a curve $H(\lambda,\varphi)$ of auxiliary Hamiltonians,
\begin{equation}\label{seq}
H(\lambda,\varphi) \mapsto \ket{\Phi(\lambda,\varphi)}  \mapsto \g(\lambda,\varphi)\,,
\end{equation}
whose ground state 1RDMs $\g(\lambda,\varphi)\equiv \mbox{Tr}_{N-1}[\ket{\Phi(\lambda,\varphi)}\!\bra{\Phi(\lambda,\varphi)}]$ describe a straight line at an angle $\varphi$ starting at $\ket{\varphi}\!\bra{\varphi}=\g(\lambda=0,\varphi)$. To calculate the functional $\Fp[\g(D,\varphi)]$ for $D\ll1$ according to \eqref{gh} we expand the Hamiltonian, $H(\lambda,\varphi)\equiv h(\lambda,\varphi)+\lambda W$ $= h^{(0)} + \lambda (h^{(1)}+ W) + \mathcal{O}(\lambda^2)$. The purpose of the higher orders of the one-particle Hamiltonian $h(\lambda,\varphi)$ is to ensure that $\gamma(\lambda,\varphi)$ remains diagonal in the basis  $\ket{\varphi},\ket{\varphi^{\perp}}$, at least up to second order in $\lambda$.
As it is shown in Appendix \ref{sec:PT}, the eigenvalue problem $H(\lambda,\varphi)\ket{\Phi(\lambda,\varphi)}=E(\lambda,\varphi)\ket{\Phi(\lambda,\varphi)}$ can systematically be solved in several orders of $\lambda$, while the enforced diagonality of $\g$ determines the required higher order terms of $h(\lambda,\varphi)$.
Comparison of the 1RDM of the ground state $\ket{\Phi(\lambda,\varphi)}$ with Eq.~(\ref{1RDMsph}) fixes $\lambda\equiv\lambda(D,N)= \sqrt{DN/2}$. Plugging all results from the perturbation theoretical calculation into Eq.~\eqref{gh} yields
(for $D \ll1$):
\begin{eqnarray}\label{approx-funct}
\FpN[\g(D,\varphi))]  &\simeq & E_N^{(0)}(\varphi) + E_N^{(1)}(\varphi)D  \\
&& - U\sin^2(\varphi) N\sqrt{N-1}\sqrt{D}\,,
\nonumber
\end{eqnarray}
where $E^{(0)}_N(\varphi) \equiv UN(N-1) [ 1 - \tfrac12 \sin^2(\varphi)\big]$, $E^{(1)}_N(\varphi) \equiv UN(N-2)(3\sin^2(\varphi)-2)$ depend on  $\varphi$ and $N$ only. The key result \eqref{approx-funct} confirms the existence of a `BEC-force' on the boundary of the domain $\mathcal{P}_p$. Indeed, we find that
\begin{equation}\label{div}
\frac{\partial \mathcal{F}^{\,(p)}_N}{\partial D}
= -(U/2)N\sqrt{N-1}  \sin^2(\varphi) D^{-1/2}  + \mathcal{O}(D^0)
\end{equation}
diverges repulsively for $D \to 0$, for all $N$ and $\varphi$ (except $\varphi=0,\pi$).

To fully appreciate the scope of the surprising finding \eqref{div}, let us recall that the functional $\FpN$ is universal.
Its form and features therefore provide insides into the ground states of \emph{all} Hamiltonians $H(h)$ \eqref{ham} \emph{simultaneously}.
To illustrate this in the Hubbard dimer, we choose an arbitrary $h$ (i.e., $t$ and $\Delta v \equiv (v_L-v_R)/2t$). The energy functional follows as $\mbox{Tr}[h\g]+ \FpN[\g]$.
%With the first two terms of $H$(Eq.~(\ref{dimer})) and the circular coordinates for $\gamma$ one obtains
%$\langle h, \gamma(D,\varphi))\rangle = -tN\big[ \sin(\varphi) +v_{+} + v_{-}\cos(\varphi)\big](1-2D)$
%with the dimensionless parameters $v_{\pm}=(v_L \pm v_R)/(2t)$.
%Using this result and Eqs.~(\ref{approx-funct}) we obtain the functional $\mathcal{E}_h(\gamma(D,\varphi))$.
Its minimization yields the corresponding ground state energy and the ground state 1RDM (described by $\varphi_0$ and $D_0 \equiv 1-N_{\mathrm{BEC}}/N$), as a function of $\Delta v, u=U/t$ and $N$.
For the number $N_{\mathrm{BEC}}$ of bosons condensed in the one-particle state $\ket{\varphi_0}$ we obtain
\begin{equation}\label{condensed-number}
N_{\mathrm{BEC}} \simeq  N\left[1- \frac{(N-1) \sin^4(\varphi_0)}{16(\sin(\varphi_0) - \Delta v \cos(\varphi_0))^2} u^2  \right] \,.
\end{equation}
The required condition of BEC, $D_0= 1-N_{\mathrm{BEC}}/N \ll 1$  implies $u \ll 1/\sqrt{N-1}$. The corresponding natural orbital $\ket{\varphi_0}$ typically deviates from the lowest eigenstate of $h$, but its concrete form is here not relevant.
%The angle $\varphi_0$ depends mainly on the ratio $[u(N-1)/v_{-}]$ of the Bose-Hubbard coupling constant and the `gradient' $(v_L - v_R)$ of the external potential. One finds
%\begin{align}\label{phi-0-a}
%\varphi_0(N;v_{-},u) \simeq \varphi_{*}(v_{-})+(N-1)\cos(\varphi_{*}(v_{-}))\frac{u}{1+v_{-}^2}
%\end{align}
%for $v_{-} \ll (N-1)u$  and
%\begin{align}\label{phi-0-b}
%\varphi_0(N;v_{-},u) \simeq \frac{\pi}{2}+(N-1)\cos(\varphi_{*}(v_{-}))\frac{v_{-}}{1+u^2} \  .
%\end{align}
%for $v_{-} \gg (N-1)u$.  The angle $\varphi_{*}(v_{-})$ in Eq.~(\ref{phi-0-a}) is the minimizer of the one-particle energy
%$\langle h, \gamma(D,\varphi))\rangle$  for  $D=0$. It follows from $\cot(\varphi_{*})=-v_{-}$.

\paragraph*{Bogoliubov-approximated systems.---}
As a second example we discuss homogeneous dilute Bose gases with an arbitrary pair interaction $W(|\vec{r}_i-\vec{r}_j|)$ in a cubic box of length $L$. We exploit the commonly used $s$-wave scattering approximation and recall that the pair interaction simplifies in the dilute regime to
$\frac{W_0}{2L^3}\hat{n}_0(\hat{n}_0-1)$ $+\frac{W_0}{2L^3}\sum_{\bd{p}\neq 0} \left(2\hat{n}_0 \hat{n}_{\bd{p}}+b_{\bd{p}}^\dagger b_{-\bd{p}}^\dagger b_0 b_0 +h.c.\right)$, where $W_0$ denotes the zeroth Fourier coefficient of $W(\cdot)$ \cite{pita}.
%This approximation is valid in the dilute regime as characterized by $n a^3 \ll 1$ where
%Bogoliubov's result for $N_{\max}(na^3)$ is exact up to first order in $\sqrt{na^3 }$ \cite{Lee1957b}.
As a consequence, the functional $\mathcal{F}[\{n_{\bd{p}}  \}_{\bd{p}\neq 0}]$ separates, $\mathcal{F}[\{n_{\bd{p}}\}] = \sum_{\bd{p}\neq 0}\mathcal{F}_{\bd{p}}[n_{\bd{p}}]$.
%In principle the functional could be calculated again by the constrained search \eqref{levy}. Yet,
Moreover, the contribution $E_{\epsilon_{\bd{p}}}$ of each pair mode $(\bd{p},-\bd{p})$ to the ground state energy is known for any choice of the kinetic energy $\sum_{\bd{p}}\epsilon_{\bd{p}}\hat{n}_{\bd{p}}$, $E_{\epsilon_{\bd{p}}}=\frac12\left[\sqrt{\epsilon_{\bd{p}}^2+2n W_0 \epsilon_{\bd{p}}} - (\epsilon_{\bd{p}}+n W_0)\right]$, where $n\equiv N/L^3$ denotes the particle density. This allows us to determine $\mathcal{F}_p[n_{\bd{p}}]$ more directly as the Legendre-Fenchel transform of $E_{\epsilon_{\bd{p}}}$ (cf.~Eq.~(\ref{energy-h}) and \cite{L83,Schilling2018}), leading to $\mathcal{F}_{\bd{p}}[n_{\bd{p}}] = E_{\epsilon_{\bd{p}}(n_{\bd{p}})} - \epsilon_{\bd{p}}(n_{\bd{p}}) n_{\bd{p}}$. $\epsilon_{\bd{p}}(n_{\bd{p}})=(n W_0/2)[(2n_{\bd{p}}+1)/\sqrt{n_{\bd{p}}(n_{\bd{p}}+1)}-2]$  follows from the inversion of the known relation $n_{\bd{p}}\equiv n_{\bd{p}}(\epsilon_{\bd{p}})$ \cite{pita}.
Eventually, this yields
%the asymptotically exact result for any dilute Bose gas,
\begin{equation}\label{functional-homo}
\FpN[\{n_{\bd{p}}\}] \simeq -n W_0\sum_{{\bd{p}}\neq 0}  \big[\sqrt{n_{\bd{p}} (n_{\bd{p}} + 1)}   - n_{\bd{p}}   \big] \,.
\end{equation}
In analogy to the dimer's result (\ref{approx-funct}), any homogeneous dilute Bose gas exhibits a `BEC force' which diverges repulsively on the boundary of $\mathcal{P}_p$. To illustrate this, we consider a straight path to the boundary $\partial \mathcal{P}_p$.
Taking the derivative of the functional \eqref{functional-homo} along that path with respect to the distance $D \equiv 1- N_\mathrm{BEC}/N$ close to complete BEC yields $\mathrm{d} \FpN/\mathrm{d} D\propto -1/\sqrt{1-N_{\mathrm{BEC}}/N}$. Hence, the diverging `BEC force' prevents the system from reaching complete BEC.
%$$D= N-N_{BEC} \equiv \sum_{\vec{p} \neq 0} n_{\vec{p}} =0$$
%we use the scaling
%$$n_{\vec{p}}= (1-N_{BEC}/N) \tilde{n}_{\vec{p}}$$
%with
%$$\frac{1}{N}\sum_{\vec{p} \neq 0} \tilde{n}_{\vec{p}}=1 $$.
%Then it follows
%$$d \mathcal{F}^{(p)}_N/ d D \simeq -(W_0 n)/2 [\sum_{\vec{p} \neq 0}
%\sqrt{\tilde{n}_{\vec{p}}}] 1/ \sqrt{D} = \mathcal{O}(1)$$

\paragraph*{Conclusion.---}  Bose-Einstein condensation (BEC) is
often described through the Gross-Pitaevskii mean-field theory \cite{Pitaevskii1961,Gross1963,pita}. We have proposed a reduced density matrix functional theory (RDMFT) which no longer discards the quantum correlations but recovers them in an exact way.
In contrast to its fermionic counterpart \cite{Schilling2018}, the underlying one-body $N$-representability problem is trivial and cannot hamper the development of bosonic RDMFT. By solving the $v$-representability problem for the building block of optical lattices ($N$-boson Hubbard dimer)
we identified Levy's constrained search as the ideal starting point for constructing accurate functional approximations.
This allowed us to determine for two classes of systems the exact functionals $\F[\g]$. Remarkably, their gradients were found to  diverge in the regime of Bose-Einstein condensation, $\nabla_\g \F \propto 1/\sqrt{1-N_{\mathrm{BEC}}/N}$, providing a natural explanation for the absence of
\emph{complete} BEC in nature. For its proof, we developed a general approach which facilitates the calculation of functionals close to the boundary of their domains. This key finding of a universal `BEC-force' can be seen as the bosonic analogue of the recently discovered fermionic exchange force \cite{Schilling2019}.

We also would like to reiterate that $\F[\gamma]$ is universal. It depends only on the interparticle interaction $W$ while the one-particle terms $h$ are covered by the linear functional $\mbox{Tr}[h \g]$. Hence, determining or approximating $\F[\gamma]$ would represent the simultaneous (partial) solution of the ground state problem for \emph{all} Hamiltonians of the form $H(h)= h+W$.  This offers a range of new possibilities. For instance, any trap potential could be considered and linear response coefficients become accessible. Furthermore, in analogy to many-body localization for electrons (see, e.g., Ref.~\cite{Basko2006} and references therein), the influence of disorder and interparticle interactions on BEC and their competition can be studied in a more direct manner. All those natural applications highlight the promising potential of bosonic RDMFT.

\begin{acknowledgements}
We thank J.M.\hspace{0.5mm}Gracia-Bond\'ia and J.\hspace{0.5mm}Schmidt for helpful discussions.
CS acknowledges financial support from the Deutsche Forschungsgemeinschaft (Grant SCHI 1476/1-1) and the UK Engineering and Physical Sciences Research Council (Grant EP/P007155/1).
\end{acknowledgements}

\bibliography{Refs2}

%\onecolumngrid
%\newpage
%\begin{center}\Large{\textbf{Supplemental Material}}
%%\Large{\textbf{``Reconstructing quantum states from single-party information''}}
%\end{center}
%\setcounter{equation}{0}
%\setcounter{figure}{0}
%\setcounter{table}{0}
%%\setcounter{page}{1}
%\makeatletter
%\renewcommand{\theequation}{S\arabic{equation}}
%\renewcommand{\thefigure}{S\arabic{figure}}
%%\renewcommand{\bibnumfmt}[1]{[S#1]}
%%\renewcommand{\citenumfont}[1]{S#1}
%\vspace{0.5cm}

\appendix

\onecolumngrid
\newpage

\section{Solution of the $v$-representability problem for the $N$-boson Hubbard dimer}\label{sec:nonV}

%\subsection{Notation}
%The universal pair interaction of the Bose-Hubbard dimer is described by the operator
%\begin{equation}\label{SMW}
%\hat{W} =  \sum_{j=L/R} \hat{n}_j(\hat{n}_j-1).
%\end{equation}
Any $N$-boson state $\ket{\Psi}$ can be expressed as a linear combination
%\begin{equation}\label{SMconf}
%\ket{\Psi} = \sum^N_{n=0} \alpha_n \ket{n,N-n}
%\end{equation}
of the configuration states
\begin{align}\label{states}
\ket{n,N-n} \equiv \frac1{\sqrt{n!(N-n)!}} (b^\dagger_L)^n(b^\dagger_R)^{N-n}\ket{0}.
\end{align}
By denoting the real-valued expansion coefficients by $\alpha_n$ the 1RDM follows as
\begin{align}
\label{eqLR}
\gamma_{LR} = \frac1{N}\sum_{n=0}^{N-1} \sqrt{(N-n)(n+1)}\, \alpha_n \alpha_{n+1}
\end{align}
and
\begin{align}
\label{eqLL}
\gamma_{LL} = \frac1{N} \sum_{n=0}^{N} n \,\alpha_n^2.
\end{align}

%\subsection{Non-$v$-representable 1RDM (`blacked filled ellipses')}
To prove that any 1RDM in the black filled ellipses (see Figure 2) is not $v$-representable, let us recall the trivial solution of the eigenvalue problem of
\eqref{dimer} for $t=0$ (zero hopping). In that case, the eigenstates of the Hamiltonian are just the configuration states $\ket{n,N-n}$ \eqref{states} with corresponding energies (setting $U\equiv 1$)
\begin{equation}
E_n = \frac{N}{2}\left(v_L+v_R\right)+\left(n-\frac{N}{2}\right) (v_L-v_R)+ n(n-1)+(N-n)(N-n-1)\,.
\end{equation}
By varying the potential energy difference $v_L-v_R$ each off the $N+1$ configuration states can be reached as a ground state. In particular, to consecutive configurations $(n,N-n)$, $(n+1,N-n-1)$ become degenerate for the specific value
\begin{equation}
\Delta v_n^\ast \equiv v_L-v_R = 2 (N-1-2n)\,.
\end{equation}
By referring to degenerate perturbation theory, this implies that any corresponding superposition
\begin{equation}\label{Psix}
\ket{\Psi} = x\ket{n,N-n} \pm \sqrt{1-x^2} \ket{n+1,N-n-1}
\end{equation}
can be reached from an infinitesimal deformation of the initial one-particle Hamiltonian in Eq.~\eqref{dimer} with $t=0$ and a potential difference $\Delta v_N^\ast$. The 1RDM of \eqref{Psix} follows as (recall Eqs.~\eqref{eqLR} and \eqref{eqLL}):
\begin{equation}\label{gx}
\gamma_{LL} = \frac1{N}(n + 1 - x^2)\,,\qquad
\gamma_{LR} = \pm\frac1{N}\sqrt{(N-n)(n+1)} \, x \sqrt{1-x^2}\,.
\end{equation}
By varying $x$ and considering both signs $\pm$ the respective family \eqref{gx} of 1RDMs give rise to an ellipse, described by
\begin{equation}\label{SMellipse}
\left[N\gamma_{LL} - \left(n + \frac12\right)\right]^2 + \frac{N^2\gamma_{LR}^2}{(n+1)(N-n)} = \frac14\,.
\end{equation}
There are in total $N$ such ellipses, with centers at $\gamma_{LL}= (2n+1)/2N, \g_{LR}=0$, $n=0,1,\ldots,N-1$.
The ellipses's minor radius is equal to $1/2N$ and the major radius follows as $\sqrt{(N-n)(n+1)}/2N$. Therefore the ellipses' areas follow as \begin{equation}
a_n = \pi \sqrt{(N-n)(n+1)}/4N^2\,.
\end{equation}
Any two neighboring ellipses `touch' in one point on the axis with $\g_{LR}=0$ (see also Figure 2).

In the following we prove the key result that any 1RDM which is surrounded by one of the $N$ ellipses is not $v$-representable (and therefore shown in `black' in Figure 2). Since all the other non-$v$-representable 1RDMs lie on the boundary of the disc $\mathcal{P}_p$ and have therefore no volume (in $\mathbb{R}^2$), the probability $p_N=1-\mbox{Vol}(\mathcal{P}_v)/\mbox{Vol}(\mathcal{P}_p)$ for finding non-$v$-representable 1RDM is given by the expression
\begin{align}
p_N = \frac{\sum^{N-1}_{n=0} \sqrt{(N-n)(n+1)}}{N^2}
\end{align}
For instance, we find $p_2 = 0.71$, $p_3 = 0.60$, $p_4 = 0.56$ and $p_N$ converges to $\pi/8\simeq 0.39>0$ in the limit $N\rightarrow \infty$.

To proceed, we recall that the minimization
\begin{equation}\label{SMtotalF}
E(h)= \min_{\g \in \mathcal{P}_p}\mathcal{E}_h[\g]\equiv \min_{\g \in \mathcal{P}_p} \big[ \mbox{Tr}[h \g] +\Fp[\g]\big]  \,
\end{equation}
of the total energy functional $\mathcal{E}_h[\cdot]$ is nothing else than the Legendre-Fenchel transform
of $\mathcal{F}_p$ (up to minus signs) \cite{Schilling2018}.
Most importantly, the right-hand side of \eqref{SMtotalF} has thus a clear geometric meaning \cite{Schilling2018}. To explain this, we observe
\begin{equation}
\mbox{Tr}[h \g] = (v_L-v_R,-2t)\cdot (\g_{LL},\g_{LR})+v_R
\end{equation}
and introduce the graph of $\Fp$,
\begin{equation}
\mbox{graph}(\Fp) \equiv \Big\{\big(\g_{LL},\g_{LR},\Fp[\g_{LL},\g_{LR}]\big)\,|\, (\g_{LL},\g_{LR})\in \mathcal{P}_p\Big\}\,\subset \,\mathbb{R}^3\,.
\end{equation}
The process of minimizing $\mathcal{E}_h[\g_{LL},\g_{LR}]$ on the space $\mathcal{P}_p \subset \mathbb{R}^2$ thus means to consider a hyperplane
in $\mathbb{R}^3$ with normal vector $(v_L-v_R,-2t,-1)$ and move it upwards (i.e., in the positive $z$-direction) until it touches the graph of $\Fp$. The intercept of that hyperplane with the $z$-axis is the ground state energy $E(h)\equiv E(v_L,v_R,t)$ and each point on the graph touching that hyperplane is a possible ground state 1RDM (not exclusively referring to pure ground states). For each Hamiltonians $H(h)$ with a nondegenerate ground state there is consequently only one such $\g$. Yet, for our $N-1$ specific one-particle Hamiltonians $h$ with $t=0$ and potential energy difference $\Delta v_n^\ast$ this is quite different. Any ellipse \eqref{SMellipse} embedded into $\mathbb{R}^3$ with the $z$-values $E(h)-\mbox{Tr}[h \g]$ are contained in the graph of $\Fp$ (shown in black in Figure 2b). Due to the geometric interpretation of the minimization \eqref{SMtotalF}, the 1RDMs surrounded by the respective ellipse can be obtained as ground state 1RDMs only for the same choice $h$, i.e., $t=0$ and $v_L-v_R = \Delta v_n^\ast$. It remains to confirm that those 1RDMs do necessarily correspond to mixed ground states. That is obvious though since all pure ground states of the Hamiltonian with $t=0$ and potential energy difference $\Delta v_n^\ast$
take the form \eqref{Psix} with 1RDMs on the ellipse rather than surrounded by it. This proves that any 1RDM in the interior of the black filled ellipses in Figure 2 is not $v$-representable.

\section{Exact functionals for the Bose-Hubbard dimer}\label{sec:Fdimer}

\subsection{Pure functional for $N=2$}

In this section we focus in the case $N = 2$ for the boson dimer.
The dimension of the Hilbert space is 3 with basis set $\{\ket{2,0},\ket{1,1},\ket{0,2}\}$
(see Eq.~\eqref{states}). A wave function belonging to such a space then reads:
\begin{align}
\label{eq2d}
\ket{\Psi} = \alpha_0 \ket{2,0} + \beta\ket{1,1} + \alpha_1 \ket{0,2},
\end{align}
with the normalization condition (say, $\beta^2 + \alpha_0^2 + \alpha_1^2 = 1$).
The corresponding 1RDM is fully determined by
the equations \eqref{eqLR} and \eqref{eqLL}, which for the case of the state \eqref{eq2d} follow as
\begin{align}
\gamma_{LL} =  \frac1{2}\left(\beta^2 + 2 \alpha_0^2\right)
\end{align}
and
\begin{align}
\gamma_{LR}  = \frac{\sqrt{2}}{2}
(\alpha_0 + \alpha_1)\beta.
\end{align}
The functional we are looking is defined according to
\begin{align}
\label{fun1}
\mathcal{F}^{(p)}[\gamma] =  \min_{\Psi \rightarrow \gamma}
 \bra{\Psi} \hat W  \ket{\Psi},
\end{align}
where $\hat{W}=\sum_{j=L/R} \hat{n}_j(\hat{n}_j-1)$ (c.f Eq.~\eqref{dimer}).
We obtain $ \bra{\Psi}\sum_i \hat W  \ket{\Psi} =
2 (2\alpha_0^2 + \beta^2 + 2 \alpha_1^2) -2 = 2( 2- \beta^2) - 2$. Moreover, we find $\alpha_0^2 - \alpha_1^2 = 2\gamma_{LL} - 1$
and $(\alpha_0 + \alpha_1)^2 = 2\gamma_{LR}^2 /\beta^2$.
Therefore,
\begin{align}
\alpha_0^2 + \alpha_1^2 = \frac{( \gamma_{LL}-\tfrac12)^2}{\gamma_{LR}^2}\beta^2 + \frac{\gamma_{LR}^2}{\beta^2},
\end{align}
which gives
$(1- \beta^2)\beta^2 /\gamma_{LR}^2 - \left(\gamma_{LL}-\tfrac12\right)^2\beta^4 - \gamma_{LR}^4 = 0$.
This last equation is an equation for $\beta^2$, whose solutions are
\begin{align}
\label{fun2}
\beta^2 = \frac{1 \pm \sqrt{1 - 4\left[\gamma_{LR}^2 + (\gamma_{LL}-\tfrac12)^2\right]}}
{2[\gamma_{LR}^2 + (\gamma_{LL}-\tfrac12)^2]} \gamma_{LR}^2.
\end{align}
Putting together Eqs.~\eqref{fun1} and \eqref{fun2} we obtain the expression for
the functional, namely:
\begin{align}\label{fun2D}
\mathcal{F}^{(p)}_2[\gamma]
= 2 -
\frac{1 + \sqrt{1 - 4\left[\gamma_{LR}^2 + (\gamma_{LL}-\tfrac12)^2\right]}}
{[\gamma_{LR}^2 + (\gamma_{LL}-\tfrac12)^2]} \gamma_{LR}^2.
\end{align}
Taking polar coordinates as introduced in the main text
$\gamma_{LL}(D,\varphi) = \tfrac12[1+(1-2D) \cos(\varphi)]$ and
$\gamma_{LR}(D,\varphi) = \tfrac12 (1-2D) \sin(\varphi)$,
\begin{align}
\label{fund}
\mathcal{F}^{(p)}_2[\gamma(D,\varphi)]
=   2 - \left(1 + \sqrt{1 - (1-2D)^2}\right) \sin^2(\varphi).
\end{align}
Since a system of 2 boson is equivalent to the singlet sector of the 2-fermion problem, the functional \eqref{fund} retains some similarities with functional for the Fermi-Hubbard dimer \cite{Cohen1}.

\subsection{Ensemble functional for large $N$}

For each 1RDM on the boundary of the allowed region, $\g(D=0,\varphi)=\ket{\varphi}\!\bra{\varphi}$, there exists only one corresponding $N$-boson quantum state mapping to $\g(0,\varphi)$, namely the state which populates the orbital $\ket{\varphi}$ with all $N$ bosons, $\ket{\Psi}= \frac{1}{\sqrt{N!}}\,(b_{\varphi}^\dagger)^N\ket{0}$.
The functional on those points yields
\begin{eqnarray}\label{ensemblefun0}
\mathcal{F}^{(e)}_N[D=0,\varphi] = \bra{\Psi}\hat W\ket{\Psi} &=&
N(N-1)\left[1 - \tfrac12 \sin^2(\varphi) \right] \nonumber \\
&=& N(N-1)(1-2\g_{LR}^2) \nonumber \\
&=& N(N-1) \left[\frac{1}{2}+2\Big(\g_{LL}-\frac{1}{2}\Big)^2\right] \,.
\end{eqnarray}
On the other hand, the ground state of $H(h)$ with $t=0$ and potential energy difference $v_L-v_R=\Delta v_n^\ast$ (recall Section \ref{sec:nonV}) is given by the configuration state $\ket{n,N-n}$, where $n=0,1,\ldots,N$. This allows us to determine the values of the ensemble (and also pure) functional for specific values on the axis characterized by $\g_{LR}=0$,
\begin{eqnarray}
\mathcal{F}^{(e)}_N[\gamma_{LL}=n/N, \gamma_{LR}= 0] &=& n^2 + (N-n)^2 -N \nonumber \\
 &=& \left(\frac{N^2}{2}-N\right)+2 \left(\g_{LL}-\frac{1}{2}\right)^2 N^2\,.
\end{eqnarray}
In the limit $N\rightarrow \infty$ this relation holds for all $\g_{LL} \in [0,1]$.
%By $n = ND$ for $n \leq N/2$, yielding
%\begin{align}
%\label{ensemblefun}
%\mathcal{F}^{(e)}_N(D, \varphi =0) = (ND)^2 + (N - ND)^2 - N
%=N^2 - 2N^2(1-D)D-N,
%\end{align}

For each $\g_{LL}$ the values of $\FeN$ at $\g_{LR}=0$ and on the boundary of the disc $\mathcal{P}_p$
coincide in leading order in $N$. Due to the convexity of the ensemble functional this implies that $\FeN$ in the limit of large $N$ is independent of $\g_{LR}$ and follows as
\begin{equation}
\lim_{N\rightarrow \infty} \left[\frac{2}{N^2}\FeN[\g_{LL},\g_{LR}]-1\right] = 4 (\gamma_{LL}-\tfrac12)^2\,.
\end{equation}
As a matter of fact, since the minor radius $1/2N$ of the black filled ellipses (see Figure 2) is getting smaller and smaller for increasing $N$, the pure and ensemble functionals coincide in the limit $N\rightarrow \infty$.

\section{Perturbational construction of the functional}\label{sec:PT}
This Section presents details of our perturbational approach to the calculation of the functional
$\mathcal{F}_N(\gamma(D,\varphi))$ for $\gamma$ which are $v$-representable. The fact that $\gamma(D,\varphi)$ is $v$-representable for every $\varphi$ (except $\varphi\neq 0,\pi$) for $D$ small enough simplifies this task. Since the functional is linear in the coupling constant $U$ of the two-body interaction of the Hamiltonian \eqref{dimer}, we put $U=1$ and reintroduce $U$ at the end. More generally, we consider a general Hamiltonian of the form $\hat{H}=\hat{h}+\hat{W}$ and denote the 1RDM of its ground state $\ket{\Phi_{\hat{H}}}$ by $\gamma_{\hat{H}}$.

Then, according to the constrained search formalism the functional at $\g_{\hat{H}}$ follows directly from the ground state energy $E(\hat{H})$, after subtracting the one-particle energy $\langle \hat{h}\rangle_{\Phi_{\hat{H}}}= N\mbox{Tr}[\g_{\hat{H}}\hat{h}]$,
\begin{equation}\label{Fvrep}
\mathcal{F}_{\hat{W}}[\g_{\hat{h}+\hat{W}}] = E(\hat{h}+\hat{W})-N\mbox{Tr}[\g_{\hat{h}+\hat{W}}\hat{h}]\,.
\end{equation}
It is crucial here to recall that the functional $\mathcal{F}_{\hat{W}}$ depends on the interaction $\hat{W}$ only and the relation \eqref{Fvrep} is valid for any choice of $\hat{h}$.
Moreover, here and in the following we suppress the superindex $p/e$ of $\mathcal{F}$ since the pure ($\mathcal{F}^{(p)}$) and ensemble functional ($\mathcal{F}^{(e)}$) coincide for $v$-representable 1RDMs $\g$, $\mathcal{F}^{(p)}[\g]=\mathcal{F}^{(e)}[\g]$.

At first sight, determining the relation $\g \leftrightarrow \hat{h}+\hat{W}\leftrightarrow E(\hat{h}+\hat{W})$ between $v$-representable 1RDMs $\g$ and the corresponding Hamiltonians $\hat{h}+\hat{W}$ seems to be impossible. Yet the restriction to almost complete Bose-Einstein condensates, $D\approx 0$, simplifies the task considerably: For each 1RDM on the boundary of the allowed region, $\g(D=0,\varphi)=\ket{\varphi}\!\bra{\varphi}$, there exists only one corresponding $N$-boson quantum state mapping to $\g(0,\varphi)$, namely the state which populates the orbital $\ket{\varphi}$ with all $N$ bosons, $\ket{\Phi}= \frac{1}{\sqrt{N!}}\,(b_{\varphi}^\dagger)^N\ket{0}$. Then, as it is worked out in the following, a modified second order perturbation theoretical approach will allow us to establish the required relation $\g \leftrightarrow \hat{h}+\hat{W}\leftrightarrow E(\hat{h}+\hat{W})$ for all $\g(D,\varphi)$ with $D \ll 1$. For this, we start with the unperturbed Hamiltonian $\hat{h}^{(0)}\equiv -\hat{n}_{\varphi}\equiv - b_\varphi^\dagger b_\varphi$, whose non-degenerate ground state is given indeed by $\ket{\Phi}= \frac{1}{\sqrt{N!}}\,(b_{\varphi}^\dagger)^N\ket{0}$ with 1RDM $\gamma = \ket{\varphi}\!\bra{\varphi}$. Adding a perturbation  $\lambda \hat{W}$ will then change in leading order the ground state but unfortunately not the distance $D$. The latter follows from the fact that $D(\lambda)$ depends analytically on $\lambda$ and $D(\lambda=0)=0$. A finite leading order $D^{(1)}\neq 0$ in $D(\lambda)=0+ \lambda D^{(1)}+ \mathcal{O}(\lambda^2)$ would then violate the constraint $D(\lambda)\geq 0$ in a neighbourhood of $\lambda=0$ (since this includes both signs for $\lambda$). This is also the reason why we need to study the effect of the perturbation including second order terms to allow $\g$ to depart from the boundary $D=0$. Quite in contrast to $D$, the angular degree $\varphi$ of the ground state 1RDM changes already in leading order and therefore the calculation of $\frac{\partial \FN}{\partial D}(D,\varphi)$ seems to be even more difficult. To circumvent these problems, we implement the second order perturbation theory in a fancier way by adding a second perturbation in the form of a one particle Hamiltonian $\hat{h}(\lambda)-\hat{h}^{(0)}\equiv \lambda\hat{h}^{(1)}+\mathcal{O}(\lambda^2)$ determined such that the perturbed 1RDM $\g(D,\varphi)$ remains diagonal in the basis $\ket{\varphi},\ket{D}$, i.e.~$\varphi(\lambda)\equiv \varphi$, at least up to second order terms.

To summarize, we study the ground state problem of the Hamiltonian
\begin{equation}\label{Hpt}
\hat{H}(\lambda) = \hat{h}(\lambda) + \lambda \hat{W} = \hat{h}^{(0)}+ \lambda (\hat{h}^{(1)}+\hat{W})+ \mathcal{O}(\lambda^2)\,,
\end{equation}
where $\hat{h}^{(0)}=-\hat{n}_{\varphi}$ and the second and higher order terms are one-particle Hamiltonians.
We expand its ground state energy $E(\lambda)$, ground state $\ket{\Phi(\lambda)}$ and ground state 1RDM $\g(\lambda)$ in $\lambda$,
\begin{eqnarray}\label{pt2nd}
E(\lambda)&\equiv& E^{(0)} + \lambda E^{(1)}+ \lambda^2 E^{(2)}+ \mathcal{O}(\lambda^3)\nonumber \\
\ket{\Phi(\lambda)}&\equiv& \ket{\Phi^{(0)}} + \lambda \ket{\Phi^{(1)}}+ \lambda^2 \ket{\Phi^{(2)}}+ \mathcal{O}(\lambda^3)\nonumber \\
\g(\lambda)&\equiv& \g^{(0)} + \lambda \g^{(1)}+ \lambda^2 \g^{(2)}+ \mathcal{O}(\lambda^3)\,.
\end{eqnarray}
Here, we choose for the moment the common normalization condition $\bra{\Phi^{(0)}}\Phi(\lambda)\rangle \equiv 1$, i.e.,  $\bra{\Phi^{(0)}}\Phi^{(k)}\rangle = 0$ for all $k\geq 1$.

The outline for the remaining derivation is the following. First, we solve the (trivial) unperturbed problem eigenvalue problem for $\hat{h}^{(0)}$. Second, we determine all (required) coefficients in \eqref{pt2nd} up to second order. Third, by referring to \eqref{Fvrep},
we determine the functional (after having renormalized $\ket{\Phi(\lambda)}$ and $\g(\lambda)$, respectively, to unity):
\begin{eqnarray}\label{Fvreplong}
\mathcal{F}_{\hat{W}}[\gamma_{\hat{h}(\lambda)+\lambda \hat{W}}] &=& \frac{1}{\lambda} \mathcal{F}_{\lambda \hat{W}}[\gamma_{\hat{h}(\lambda)+\lambda \hat{W}}] \nonumber\\
&=& \frac{1}{\lambda} \left[E(\hat{h}(\lambda)+\lambda \hat{W})- N \mbox{Tr}[\gamma_{\hat{h}(\lambda)+\lambda \hat{W}} \hat{h}(\lambda)]\right] \nonumber\\
&=& \frac{1}{\lambda} \Big[E^{(0)} + \lambda E^{(1)}+ \lambda^2 E^{(2)} - N \mbox{Tr}[\g^0 \hat{h}^{(0)}]-\lambda N \left(\mbox{Tr}[\g^{(1)}\hat{h}^{(0)}]+\mbox{Tr}[\g^{(0)}\hat{h}^{(1)}]\right)  \nonumber \\
&& -\lambda^2 N \left(\mbox{Tr}[\g^{(2)}\hat{h}^{(0)}]+\mbox{Tr}[\g^{(1)}\hat{h}^{(1)}]+\mbox{Tr}[\g^{(0)}\hat{h}^{(2)}]\right) + \mathcal{O}(\lambda^3)\Big] \nonumber \\
&=& E^{(1)}- N \mbox{Tr}[\g^{(0)}\hat{h}^{(1)}] + \lambda \left[ E^{(2)} - N \left(\mbox{Tr}[\g^{(2)}\hat{h}^{(0)}]+\mbox{Tr}[\g^{(0)}\hat{h}^{(2)}]\right)\right]+ \mathcal{O}(\lambda^2)\,.
\end{eqnarray}
In the last line, we have used that the zeroth order terms cancel out and that our perturbation is designed such that $\g^{(1)}\equiv 0$. Below, we will see that also $\mbox{Tr}[\g^{(0)}\hat{h}^{(1)}]$ vanishes and that $\mbox{Tr}[\g^{(0)}\hat{h}^{(2)}]$ cancels out with a respective part of the second order term $E^{(2)}$.

The eigenstates of the unperturbed Hamiltonian $\hat{h}^{(0)}\equiv -\hat{n}_{\varphi}$ are given by
\begin{eqnarray}
\ket{n} &\equiv& \ket{n_\varphi=n, n_{\varphi_{\perp}}=N-n} \\
&\equiv& \frac{1}{\sqrt{n! (N-n)!}}\,(b_{\varphi}^\dagger)^n (b_{\varphi_{\perp}}^\dagger)^{N-n}\ket{0}\,. \nonumber
\end{eqnarray}
All those $N+1$ states are non-degenerate with corresponding energies
\begin{equation}
  E_n \equiv -\bra{n}\hat{n}_{\varphi} \ket{n} = -n\,,\quad n=0,1,\ldots,N\,.
\end{equation}
The ground state thus corresponds to $n=N$, $\ket{\Phi^{(0)}}= \ket{N}$.
To work out the second order perturbation theory, we need to determine the expressions $\bra{n}\hat{W}\ket{N}$ for all $n$. Since $\hat{W}$ is a two-particle operator, those matrix elements vanish for $n<N-2$ and we therefore need to determine it only for $n=N,N-1,N-2$.
Nonetheless, let us first consider an arbitrary $n$.  Since $\hat{W} = \sum_{j=L,R} \hat{n}_j(\hat{n}_j - 1) = \sum_{j=L,R} \hat{n}^2_j  -NU $  for $N$ bosons we have to calculate $\bra{n}(\hat{n}^2_L + \hat{n}^2_R)\ket{N}$.
Using
$b_{L}^\dagger = \alpha b_{\varphi}^\dagger +\beta b_{\varphi_{\perp}}^\dagger$,  $b_{R}^\dagger = \beta b_{\varphi}^\dagger -\alpha b_{\varphi_{\perp}}^\dagger $ with $\alpha = \cos{(\varphi/2)},\beta = \sin{(\varphi/2)}$ and similar for $b_{L}$ and $b_{R}$ it follows

\begin{equation}\label{W}
\hat{n}^2_L + \hat{n}^2_R= \hat{W}_0 + \hat{W}_1 + \hat{W}_2 \  ,
\end{equation}
with

\begin{eqnarray}\label{W012}
 \hat{W}_0 &=& (\alpha^4+\beta^4)(\hat{n}^2_{\varphi}+ \hat{n}^2_{\varphi_{\perp}}) +
  2 \alpha^2 \beta^2  \big[4 \hat{n}_{\varphi} \hat{n}_{\varphi_{\perp}} +(\hat{n}_{\varphi} + \hat{n}_{\varphi_{\perp}}) \big] \nonumber \\
\hat{W}_1 &=& (\alpha^3 \beta - \alpha \beta^3)\big[(\hat{n}_{\varphi} - \hat{n}_{\varphi_{\perp}})
(b_{\varphi}^\dagger b_{\varphi_{\perp}} + b_{\varphi_{\perp}}^\dagger b_{\varphi}) + (b_{\varphi}^\dagger b_{\varphi_{\perp}}+ b_{\varphi_{\perp}}^\dagger b_{\varphi})(\hat{n}_{\varphi} - \hat{n}_{\varphi_{\perp}}) \big] \nonumber \\
\hat{W}_2 &=& 2\alpha^2 \beta^2 \big[(b_{\varphi}^\dagger)^2 (b_{\varphi_{\perp}})^2 + (b_{\varphi_{\perp}}^\dagger)^2 (b_{\varphi})^2 \big]
 \end{eqnarray}

Since the unperturbed eigenstates $\ket{n}$ are eigenstates of $\hat{n}_{\varphi}$ and $\hat{n}_{\varphi_{\perp}}$ with eigenvalues $n$ and $N-n$, respectively, it follows

\begin{equation}\label{averageW}
\bra{n}\hat{n}^2_L + \hat{n}^2_R\ket{N}= W_0 \delta_{n,N} + W_1 \delta_{n,N-1} + W_2  \delta_{n,N-2} \  ,
\end{equation}
with
\begin{eqnarray}\label{averageW012}
 W_0 &=& (\alpha^4+\beta^4)N^2+
  2 \alpha^2 \beta^2 N  \nonumber \\
W_1 &=& 2\alpha \beta (\alpha^2 -\beta^2) (N-1)\nonumber \\
W_2 &=& 2 \sqrt{2}\alpha^2 \beta^2 \sqrt{N(N-1)}
 \end{eqnarray}

We proceed now to calculate various required terms in \eqref{pt2nd}. For this we actually need to first determine $\hat{h}^{(1)}$.
Since $\g^{(1)}= \mbox{Tr}_{N-1}[\ket{\Phi^{(0)}}\!\bra{\Phi^{(1)}}]+h.c.$, we determine $\ket{\Phi^{(1)}}$,
\begin{equation}
  \ket{\Phi^{(1)}}= -\sum_{n=0}^{N-1}\frac{\bra{n}\hat{h}^{(1)}+\hat{W}\ket{N}}{N-n}\,\ket{n}\,.
\end{equation}
Again, since $\hat{W}$ is a two particle operator, this sum restricts to $n=N-1,N-2$. Furthermore, only the term $\ket{n=N-1}$ can contribute to $\g^{(1)}$ since (in contrast to $\ket{N-2}$) it does not differ from $\ket{\Phi^{(0)}}= \ket{N}$ in more than one orbital. Consequently, $\hat{h}^{(1)}$ is determined by
\begin{equation}\label{h1}
  \bra{N-1}\hat{h}^{(1)}+\hat{W}\ket{N} =0
\end{equation}
and can be chosen as (recall \eqref{averageW})
\begin{equation}
\hat{h}^{(1)} = -\frac{W_1}{\sqrt{N}}\, \left(b_{\varphi_{\perp}}^\dagger b_{\varphi}+b_{\varphi}^\dagger b_{\varphi_{\perp}}\right)\,.
\end{equation}
Actually, in a similar (but lengthier) way we could determine $\hat{h}^{(2)}$ which shall ensure that also the second order correction $\g^{(2)}$ remains diagonal. Yet, the form of $\hat{h}^{(2)}$ turns out to be irrelevant and in particular its contribution within \eqref{Fvreplong} will cancel out since $\hat{h}^{(2)}$ is a one-particle operator.

After having determined the explicit form of the Hamiltonian \eqref{Hpt}, we perform now the perturbation theory.
Just to recall, in zeroth order, we have
$E^{(0)}= -N = N\mbox{Tr}[\g^{(0)}\hat{h}^{(0)}]$ and $\ket{\Phi^{(0)}}=\ket{N}$.
In first order, we obtain (using $\bra{N}\hat{h}^{(1)}\ket{N}=0$)
\begin{equation}
  E^{(1)}= \bra{N}\hat{W}\ket{N} = W_0
\end{equation}
and
\begin{equation}
\ket{\Phi^{(1)}} = - \frac{\bra{N-2}\hat{W}\ket{N}}{2}\,\ket{N-2}= -\frac{W_2}{2}\,\ket{N-2}\,.
\end{equation}
The second order of the energy follows as
\begin{eqnarray}
  E^{(2)}&=& - \sum_{n=0}^{N-1}\frac{\big|\!\bra{n}\hat{h}^{(1)}+\hat{W}\ket{N}\!\big|^2}{N-n} + \bra{N}\hat{h}^{(2)}\ket{N}\nonumber \\
  &=&- \frac{\big|\!\bra{N-2}\hat{h}^{(1)}+\hat{W}\ket{N}\!\big|^2}{2} + \bra{N}\hat{h}^{(2)}\ket{N} \nonumber \\
  &=&- \frac{W_2^2}{2} + \bra{N}\hat{h}^{(2)}\ket{N}\,.
\end{eqnarray}
In the second line we have used \eqref{h1} and that $\hat{W}$ is a two-body operator. The term $\bra{N}\hat{h}^{(2)}\ket{N}$ does not need to be determined since it will cancel out in \eqref{Fvreplong}. Due to the normalization condition $\bra{\Phi^{(0)}}\Phi^{(k)}\rangle = 0$ for all $k\geq 1$, $\ket{\Phi^{(2)}}$ has no contribution proportional to $\ket{N}$. Consequently, it cannot contribute to the diagonal entries of $\g^{(2)}$ which follow as
\begin{equation}\label{g2pre}
\g^{(2)} = \mbox{Tr}_{N-1}[\ket{\Phi^{(1)}}\!\bra{\Phi^{(1)}}]+\mbox{Tr}_{N-1}[\ket{\Phi^{(0)}}\!\bra{\Phi^{(2)}}+h.c.]\,.
\end{equation}
Since the second order term $\hat{h}^{(2)}$ is chosen such that $\g^{(2)}$ is still diagonal, we even have
\begin{eqnarray}\label{g2}
\g^{(2)} &=& \mbox{Tr}_{N-1}\big[\ket{\Phi^{(1)}}\!\bra{\Phi^{(1)}}\big] \nonumber \\
&=& \frac{W_2^2}{4}\mbox{Tr}_{N-1}\big[\ket{N-2}\!\bra{N-2}\big] \nonumber \\
&=& \frac{W_2^2}{4N} \left[(N-2)\,\ket{\varphi}\!\bra{\varphi}+2 \,\ket{\varphi_{\perp}}\!\bra{\varphi_{\perp}}\right]\,.
\end{eqnarray}
Consequently, $\ket{\Phi^{(2)}}$'s contribution to $\g(\lambda,\varphi)$ is of negligible order, $\mathcal{O}(\lambda^3)$, and therefore irrelevant for our purpose.

Wrapping up various results of the second order perturbation theory and reintroducing $U$ leads to
\begin{eqnarray}
\mathcal{F}_{\hat{W}}[\gamma_{\hat{h}(\lambda)+\lambda \hat{W}}]
&=& U\big\{ E^{(1)}+ \lambda \big[ - \frac{W_2^2}{2}  - N \mbox{Tr}[\g^{(2)}\hat{h}^{(0)}]\big]+ \mathcal{O}(\lambda^2) \big\}\nonumber \\
&=& U \big\{E^{(1)}+ \lambda \big[ - \frac{W_2^2}{2}  + N \bra{\varphi}\g^{(2)}\ket{\varphi}\big]+ \mathcal{O}(\lambda^2)\big\} \nonumber
\end{eqnarray}
and the (correctly normalized) 1RDM reads
\begin{equation}
\g(\lambda) = (1-\kappa_N \lambda^2)\,\ket{\varphi}\!\bra{\varphi} + \kappa_N\lambda^2\ket{\varphi_{\perp}}\!\bra{\varphi_{\perp}}+ \mathcal{O}(\lambda^3)\,,
\end{equation}
where
\begin{equation}
\kappa_N \equiv 4 (N-1) \alpha^4 \beta^4= \frac{W_2^2}{2N}\,.
\end{equation}
This allows us to identify
\begin{equation}
D(\lambda)= \kappa_N \lambda^2+ \mathcal{O}(\lambda^{(3)})
\end{equation}
implying
\begin{equation}
  \lambda = \frac{\sqrt{D}}{\sqrt{\kappa_N}}+ \mathcal{O}(D^1)\,.
\end{equation}
Finally, this leads to (plugging in $\alpha\equiv \cos{(\varphi/2)}$, $\beta\equiv \sin{(\varphi/2)}$)
\begin{eqnarray}
 \FN[D,\varphi] &=&U \big\{ E^{(1)}(\varphi)+ \frac{\sqrt{D}}{\sqrt{\kappa_N(\varphi)}}\left[ - \frac{(W_2(\varphi))^2}{2}  - N \kappa_N(\varphi)\right]+  \mathcal{O}(D^1) \big\} \nonumber \\
 &=&U \big\{ E^{(1)}(\varphi)- 2N \sqrt{\kappa_N(\varphi)} \sqrt{D} +  \mathcal{O}(D^1) \big\}\nonumber \\
 &=&U \big\{ E^{(1)}(\varphi)- N\sqrt{N-1}\sin^2(\varphi)\,\sqrt{D} +  \mathcal{O}(D^1) \big\}
\end{eqnarray}
and eventually
\begin{equation}
  \frac{\partial\FN}{\partial D}[D,\varphi] = - \frac{N\sqrt{N-1}\sin^2(\varphi)}{2\sqrt{D}} +  \mathcal{O}(D^0)\,.
\end{equation}

\end{document}